\documentclass[twocolumn,showpacs,superscriptaddress,amsmath,amssymb,prb]{revtex4}

\usepackage{graphicx}%
\usepackage{dcolumn}%

\begin{document}

\title{Evidence for CuO conducting band splitting in the nodal direction of Bi$_2$Sr$_2$CaCu$_2$O$_{8+\delta}$}

\author{A. A. Kordyuk}
\affiliation{Leibniz-Institut f\"ur Festk\"orper- und Werkstoffforschung Dresden, P.O.Box 270016, 01171 Dresden, Germany}
\affiliation{Institute of Metal Physics of National Academy of Sciences of Ukraine, 03142 Kyiv, Ukraine}

\author{S. V. Borisenko}
\affiliation{Leibniz-Institut f\"ur Festk\"orper- und Werkstoffforschung Dresden, P.O.Box 270016, 01171 Dresden, Germany}

\author{A. N. Yaresko}
\affiliation{Max-Planck-Institut f\"ur Physik Komplexer Systeme Dresden, 01187 Dresden, Germany}

\author{S.-L. Drechsler}
\affiliation{Leibniz-Institut f\"ur Festk\"orper- und Werkstoffforschung Dresden, P.O.Box 270016, 01171 Dresden, Germany}

\author{H. Rosner}
\affiliation{Max-Planck-Institut f\"ur Chemische Physik fester Stoffe, 01187 Dresden, Germany}

\author{T. K. Kim}
\author{A. Koitzsch}
\author{K.~A.~Nenkov}
\author{M. Knupfer}
\author{J. Fink}
\affiliation{Leibniz-Institut f\"ur Festk\"orper- und Werkstoffforschung Dresden, P.O.Box 270016, 01171 Dresden, Germany}

\author{R. Follath}
\affiliation{BESSY GmbH, Albert-Einstein-Strasse 15, 12489 Berlin, Germany}

\author{H. Berger}
\affiliation{Institut de Physique de la Mati\'ere Complexe, EPFL, CH-1015 Lausanne, Switzerland}

\author{B. Keimer}
\address{Max-Planck Institut f\"ur Festk\"orperforschung, D-70569 Stuttgart, Germany}

\author{S. Ono}
\author{Yoichi Ando}
\affiliation{Central Research Institute of Electric Power Industry, Komae, Tokyo 201-8511, Japan}

\date{October 30, 2003}%

\begin{abstract}
Using angle-resolved photoemission spectroscopy with ultimate momentum resolution we have explicitly resolved the bilayer splitting in the nodal direction of Bi-2212. The splitting is observed in a wide doping range and, within the experimental uncertainty, its size does not depend on doping. The value of splitting derived from the experiment is in good agreement with that from band structure calculations which implies the absence of any electronic confinement  to single planes within bilayers of Bi-2212. Other consequences of this finding are also discussed.
\end{abstract}


\preprint{\textit{xxx}}

\maketitle

Joys and pitfalls in the development of an appropriate theory for high temperature superconductors (HTSC) are intimately related with results from angle-resolved photoemission spectroscopy (ARPES), which is a direct probe of the quasi-particles and their interaction \cite{DamascelliRMP03}. Continuous improvement of the spectrometers leads not only to discovering new phenomena but also, and quite often, to a change of established paradigms. A distinguishing feature of modern ARPES is the ability to resolve the bilayer splitting (BS) of the CuO conduction band in the bilayer cuprates. For the first time such a splitting has been observed for overdoped Bi$_2$Sr$_2$CaCu$_2$O$_{8+\delta}$ (Bi-2212) \cite{FengPRL01,ChuangPRL01} and then also for optimally doped and underdoped samples \cite{ChuangXXX, KordyukPRB2002} (clearly resolved below \cite{KordyukPRB2002, BorisenkoPRB2002} and above \cite{BorisenkoPRL2003} the superconducting transition). It has been found \cite{FengPRL01, ChuangPRL01} that the observed splitting can be approximated by a momentum dependence $t_{\perp} (\cos k_x - \cos k_y)^2 /2$, which is expected for an inter-plane hopping between two CuO layers (where $t_{\perp}$ describes the interlayer hopping mainly mediated via Cu$4s$ orbitals when the splitting along the node is much less than the maximum splitting at the saddle-point) \cite{AndersenJPCS95}. 

One of the main conclusions coming from the observation of the BS is that the HTSC cuprates are not so unusual as it was assumed before. The strong correlations in these systems produce rather weak effects on the lineshape of the photoemission spectra \cite{BorisenkoPRL2003, KordyukPRL2002}, which can be described within the quasiparticle self-energy concept \cite{KimPRL2003}. They do not cause a principal modification of the electronic structure as the initially proposed electronic confinement to single planes within a bilayer can do \cite{Anderson1997}. In the next step, to address a question like "Is there still some space for strong correlation effects which are beyond the local density approximation (LDA) \cite{Drechsler} and could not be treated in terms of a renormalization?", the values of the BS should be compared between theory and experiment. Although the largest BS can be found in the spectra from the antinodal region [around the ($\pi$,0)-point], it is not a trivial task to extract its bare value from them. There are two main reasons for this: (i) both the deviations of the renormalized band positions from the positions of corresponding peaks in energy distribution curves (EDCs) and the deviations of the bare band positions from the renormalized ones depend on the self-energy vs.~frequency model \cite{KordyukPRL2002}; (ii) the superconducting gap and the pseudogap additionally complicate the analysis \cite{KordyukPRB2003}. So, at this stage, it is not clear whether the observed splitting values can be completely reconciled with a quasiparticle dressing of bare electrons. In addition at the antinodal points there is some uncertainty caused by the unresolved presence or shifts \cite{wang,xu} and unknown details of BiO derived states predicted by the LDA.

In this paper we focus on the nodal region, where (i) within the energy scale of the splitting the renormalization with binding energy can be considered as linear and (ii) the $d$-wave gap vanishes. We have found that the splitting along the nodal direction of bilayer Bi-cuprates is not zero but persists from underdoped to overdoped doping levels and for different composition. We show that the observed splitting is in good agreement with LDA based band structure calculations and is caused by the vertical O2$p_\sigma$-O2$p_\sigma$ hopping between two \textit{adjacent} CuO layers. This finding, leaving no space for the mentioned electronic confinement, puts some restrictions on possible microscopic mechanisms of high-$T_c$ superconductivity.

\begin{figure*}[t]
\includegraphics[width=17cm]{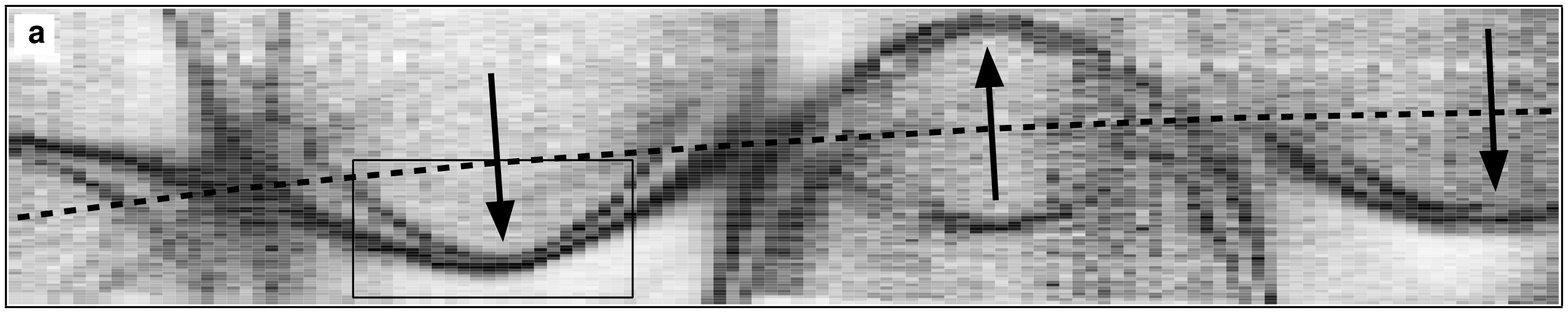}\\%
\includegraphics[width=5.6cm]{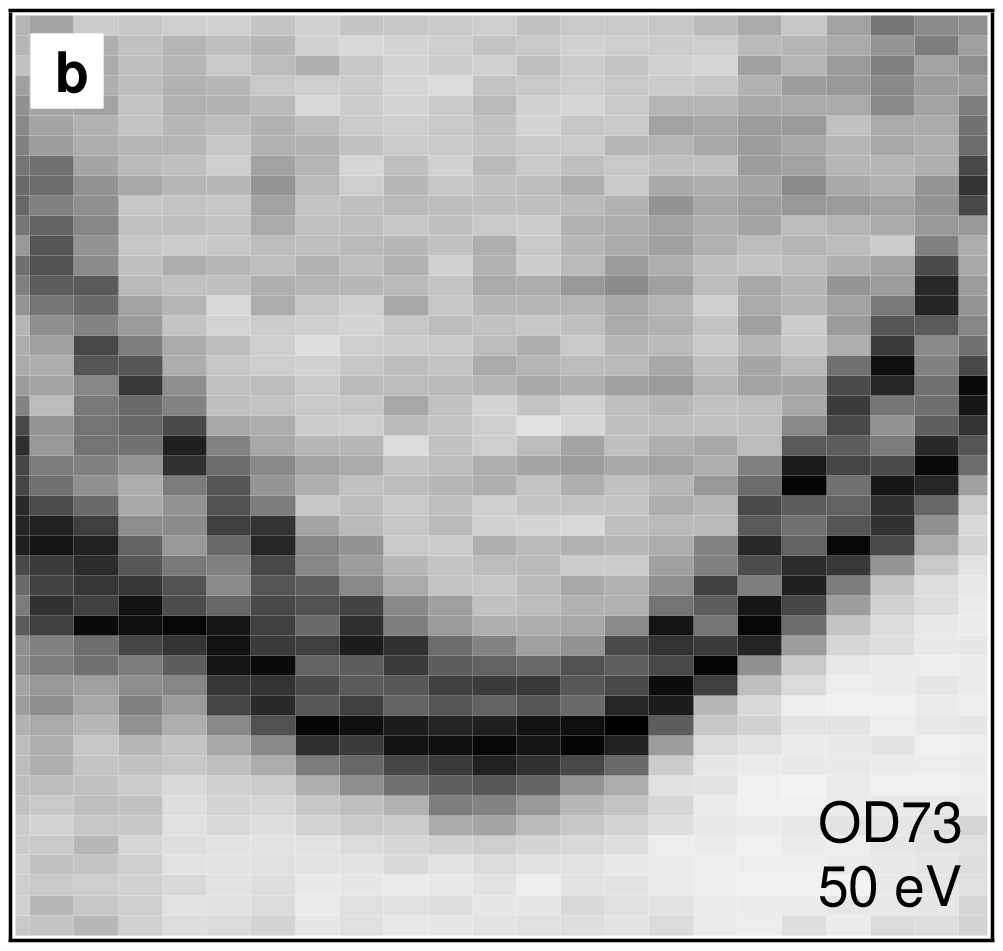}%
\includegraphics[width=5.2cm]{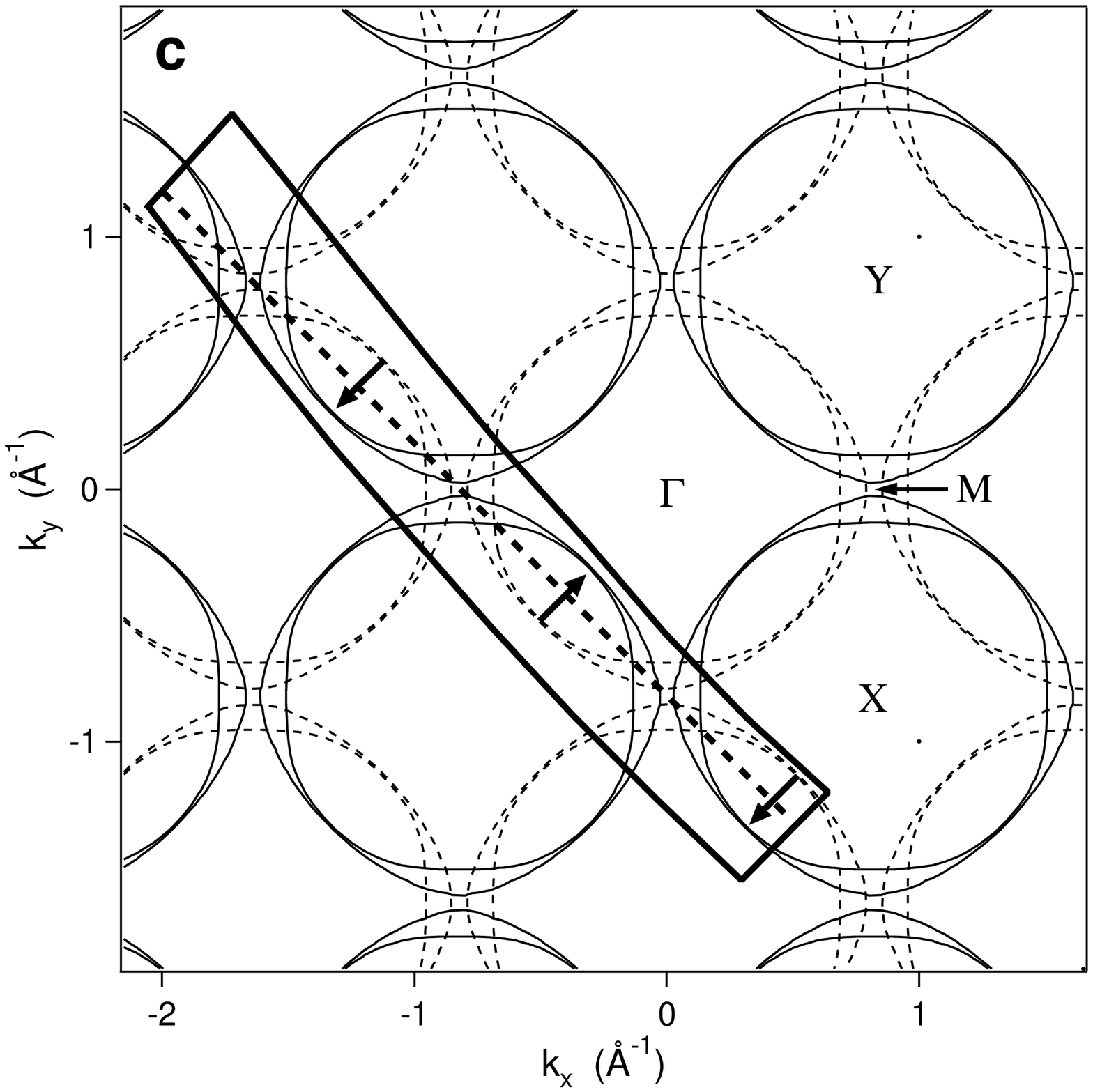}  %
\includegraphics[width=6cm]{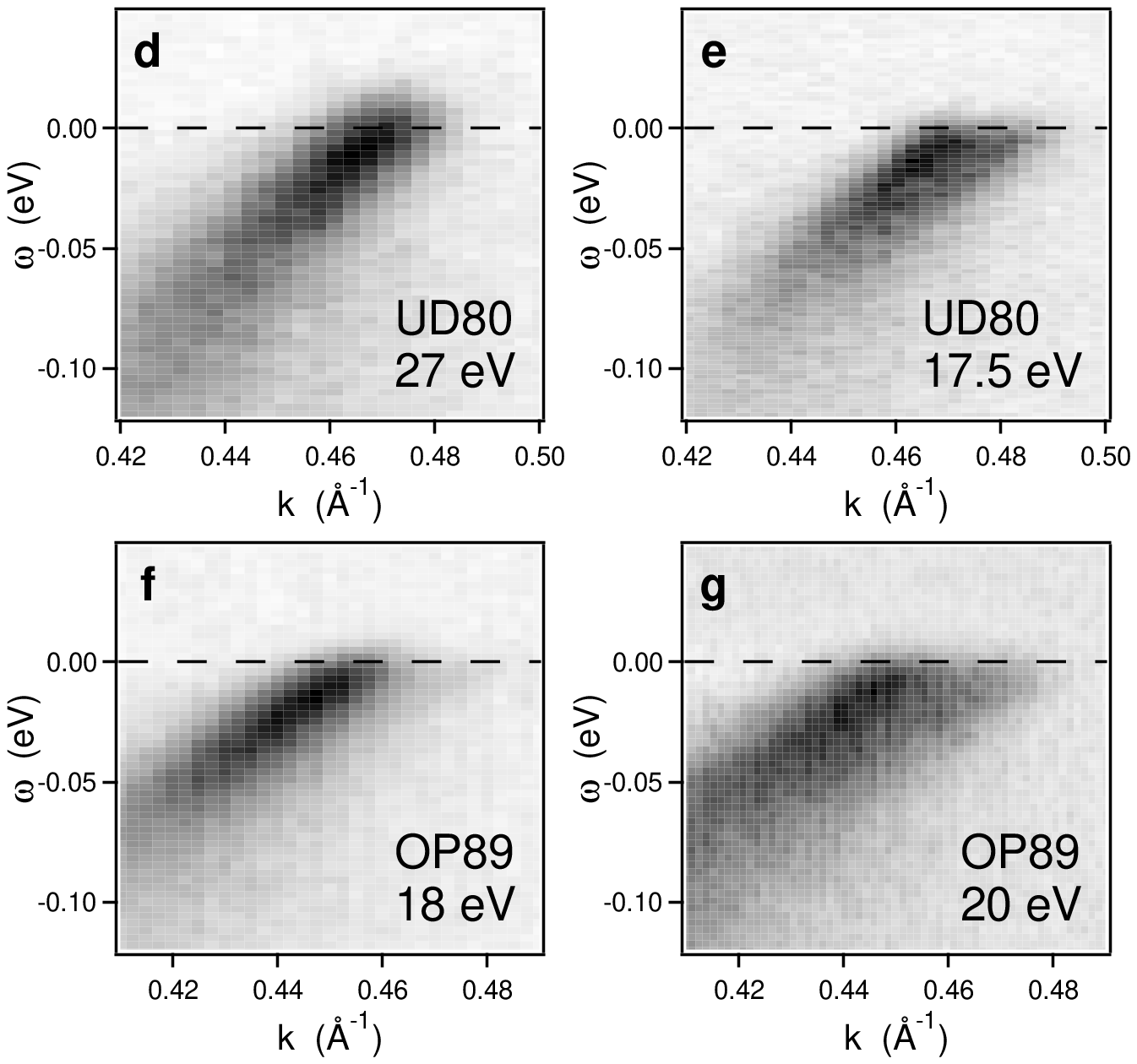}%
\caption{\label{Images} Experimental evidence for nodal splitting. (a) Fermi surface map of Bi(Pb)-2212 OD73 measured with 50 eV excitation energy at 25 K over a wide momentum region shown in (c): pointing up and down arrows mark the nodal directions in the 1st and 2nd Brilouin zone (BZ) respectively; the area of interest within the rectangle in (a) is zoomed in in (b); the dashed line goes through M-points representing the boundary of the "magnetic" BZ. (d-g) A set of images of the energy distribution maps (EDMs) for Bi-2212 UD80 (d,e) and Bi-2212 OP89 (f,g) measured at 25 K along different nodal directions at different excitation energy: (d) and (e) in the 1st BZ along the $\Gamma$Y direction, (f) and (g) in the 2nd BZ along the ZX direction.}
\end{figure*}

\begin{figure}[tb]
\includegraphics[width=2.8cm]{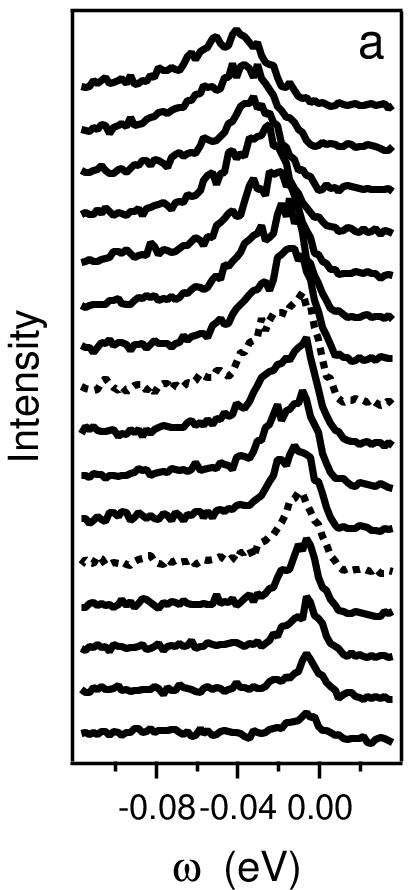}%
\includegraphics[width=2.8cm]{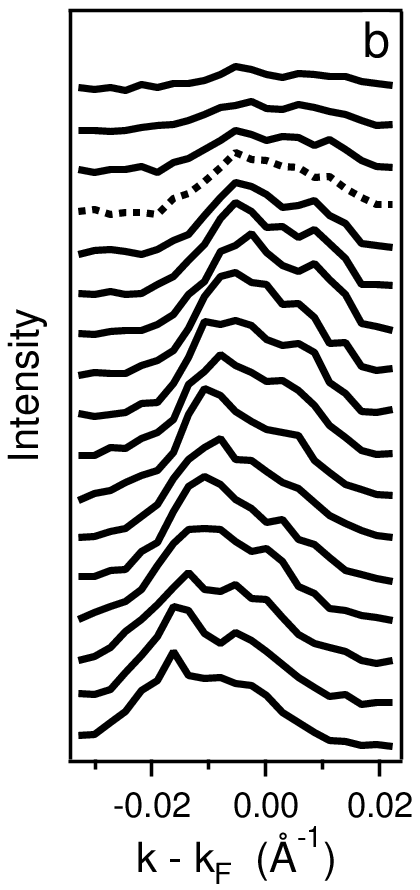}%
\includegraphics[width=2.8cm]{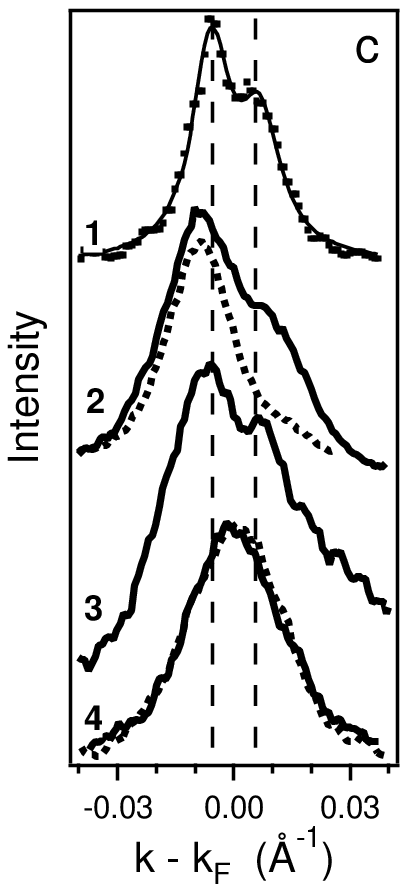}%
\caption{\label{Curves} The experimental data for Bi-2212 UD80 from Fig.~\ref{Images}e presented in form of energy distribution curves (EDCs) (a) and momentum distribution curves (MDCs) (b). EDCs are taken in the momentum range from $k_F -$ 0.025~\AA$^{-1}$ (top) to $k_F$ + 0.015~\AA$^{-1}$ (bottom), where $k_F$ is an average between antibonding, $k^a_F$, and bonding, $k^b_F$, Fermi level crossings; dotted EDCs roughly correspond to $k^a_F$ and $k^b_F$. MDCs are taken in the energy range from 3 meV (top) to $-$27 meV (bottom); $E_F$-MDC shown as a dotted curve. (c) MDCs integrated in energy about 10 meV from $E_F$: (1) -- Bi-2212 UD80, 1st BZ, 17.5 eV (solid curve represents fitting result); (2) -- Bi-2212 OP89, 2nd BZ, 20 eV for the solid curve and 18 eV for the dotted curve; (3) -- Bi(Pb)-2212 OD73, 1st BZ, 17.5 eV; (4) -- Bi-2201, 1st BZ, 17.5 eV (solid curve) and 27 eV (dotted curve).}
\end{figure}

The data have been obtained in the experimental setup where we combined a high resolution light source of a wide excitation energy range (U125/1-PGM beamline at BESSY \cite{Follath2001}), an angle-multiplexing photoemission spectrometer (SES100) and a 3-axis rotation cryo manipulator. The total energy resolution was set to 10 meV, the angular resolution of the analyzer is 0.15$^{\circ}$. The data were collected at 25 K on a bilayer superstructure-free led-doped Bi(Pb)-2212 underdoped by oxygen reduction to $T_c$ = 76 K (Pb-UD76), a pure underdoped Bi-2212 (UD80), an optimally doped Bi-2212 (OP89), an overdoped Bi(Pb)-2212 (Pb-OD73), and a single-layer Bi(La)-2201 with $T_c$ = 32 K. 

Fig.~\ref{Images} gives the experimental evidence for the nodal splitting. The Fermi surface map (normalized to maximum intensity \cite{BorisenkoPRB2001}) of Bi(Pb)-2212 OD73 is shown on panel (a). The map is measured with 50 eV excitation energy over a wide momentum region marked in (c): pointing up and down arrows mark the nodal directions in the 1st and 2nd Brilouin zones (BZs) respectively; the dashed line goes through M-points representing the boundary of the "magnetic" BZ. It is seen that the bilayer splitting is highly anysotropic but does not vanishe along the nodes. The nodal region within the rectangle in (a) is zoomed in in (b). Sketch (c) is based on the tight binding fit of the FS of an OD sample \cite{KordyukPRB2003}.

In order to demonstrate the persistence of the nodal spliting for different dopping levels we focus on the energy distribution maps (EDMs) \cite{BorisenkoPRB2001} in which the photocurrent intensity of outgoing electrons is plotted as a function of their energy and inplane momentum. We use different excitation energies that has appeared to be a powerful tool to distinguish the bilayer splitting effect on the photoemission spectra \cite{KordyukPRL2002}. Panels (d-g) show EDMs for Bi-2212 UD80 (d,e) and Bi-2212 OP89 (f,g) measured along different nodal directions at different excitation energy: (d) and (e) in the 1st BZ along the (0,0)-($\pi$,$\pi$) (or $\Gamma$Y) direction, (f) and (g) in the 2nd BZ  along the (2$\pi$,0)-($\pi$,$-\pi$) (or ZX) direction. These are the directions along which the 5x1 superstructure replicas for Pb-free Bi-2212 are well spaced \cite{KordyukPRL2002} and, therefore, do not effect the spectra in question. It is seen that for the 1st BZ the photoemission from the bonding band is suppressed at 27 eV (d) but becomes visible at 17.5 eV (e) (the data for the OP sample are similar and are not shown). In the 2nd BZ the photoemission from the bonding band is suppressed at 18 eV (f) and becomes visible at 20 eV (g). 

Fig.~\ref{Curves}(a) represents the energy cuts of Fig.~\ref{Images}(e) at constant momentum, EDCs, and (b) momentum cuts at constant energy, momentum distribution curves (MDCs). Although the photocurrent intensity at 17.5 eV is rather low, one can notice the presence of two bands in EDCs that appears as a peak with a shoulder (see EDCs in between two dotted ones in panel a). More explicitly the splitting is seen on MDCs, where two peaks can be clearly distinguished (see MDCs which are close to the $E_F$-MDC shown by the dotted line in panel b). In Fig~\ref{Curves}(c), in order to improve statistics, we integrate the MDCs along the experimental (renormalized) dispersion in the energy range 10--20 meV around $E_F$, where the MDC width does not vary dramatically. In all measured bilayer samples from UD76 to OD73 two peaks in nodal MDCs are well resolved at certain conditions which we describe below.

Exploring a wide excitation energy range (17--50 eV), we can conclude that the dependence of matrix elements on excitation energy for the nodal point in the 1st BZ exhibits a local maximum at about 17.5 eV for both the total intensity from bilayer split band and the intensity from the bonding band compared to its antibonding counterpart \cite{bonding}. The observed excitation energy dependence of the effect is in accord with recent calculations of ARPES matrix elements \cite{SahrakorpiPRB2003} which show that in the low energy range the emissions are dominated (peaked at about 18 eV) by excitation from just the O sites. For the nodal point in the 2nd BZ the dependence on matrix elements is different and the bonding band is the most pronounced for $h\nu$ = 20--21 eV excitation energy: the MDCs 2 in Fig.~\ref{Curves}(c) are derived from Fig.~\ref{Images}(f,g) and show how the bonding band peak appears when going from 18 eV (dotted curve) to 20 eV (solid curve).

In contrast to bilayer compounds, the single-layer sample shows no signature of the splitting. The nodal $E_F$ MDCs for Bi-2201 measured in the 1st BZ are presented in Fig.~\ref{Curves}c (curve 4) for two excitation energies: 17.5 eV (solid curve) and 27 eV (dotted curve). While the width of given MDCs is larger than that of each band for the bi-layer samples (due to worse surface flateness or larger scattering on impurities), its lineshape remains symmetric and excitation energy independent.  

\begin{figure}[!t]
\includegraphics[width=4.25cm]{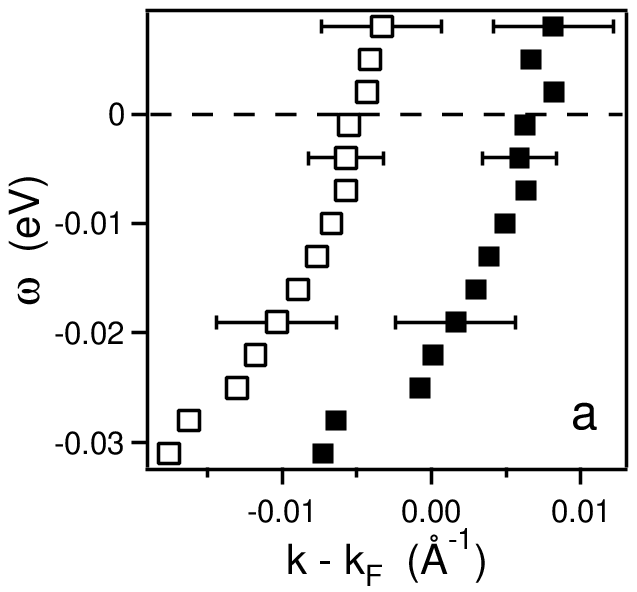}%
\includegraphics[width=4.25cm]{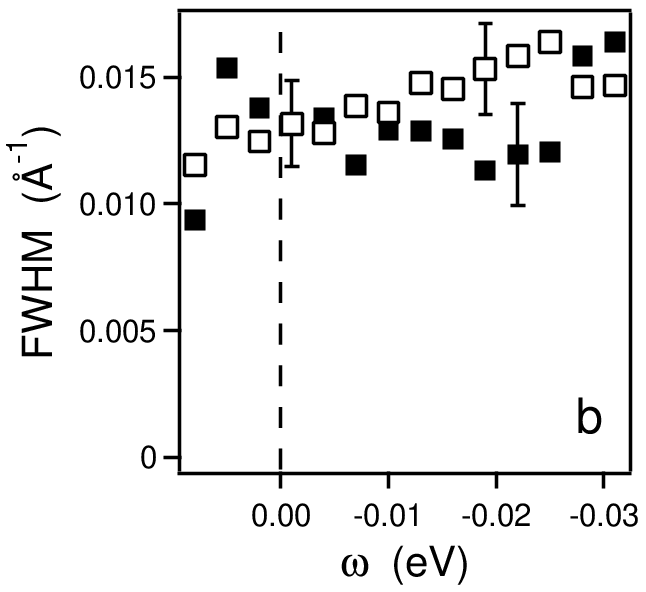}%
\caption{\label{Disp} Parameters of antibonding ($\square$) and bonding ($\blacksquare$) quasiparticle bands, dispersion (a) and width (FWHM) (b), obtained from fitting the MDCs of Fig.~\ref{Curves}b by two Lorenzians.}
\end{figure}

In order to extract precise BS values we fit the integrated MDCs to a superposition of two independent Lorenzians (an example of a fitting curve is shown in Fig~\ref{Curves}c). For the Bi-2212 UD80 sample $\Delta k$ = 0.012(1)~\AA$^{-1}$ which corresponds to 48(4) meV bare band splitting (for bare Fermi velocity $v_F$ = 4.0~eV\AA ~\cite{KordyukPRB2003}) or 23 meV splitting of the renormalized band (renormalized Fermi velocity $v^R_F$ = 2.0~eV\AA). For other bilayer samples the values of $\Delta k$ are similar: 0.015(1)~\AA$^{-1}$ (Pb-UD76), 0.015(2)~\AA$^{-1}$ (OP89), 0.014(3)~\AA$^{-1}$ (Pb-OD73).

To compare the experimentally derived BS value to LDA predictions we use two different bandstructure calculational codes: LMTO \cite{Yaresko} and FPLO \cite{Rosner}. In particular, there is a perfect agreement with LMTO calculations, according to which the BS in the nodal direction is 50 meV at $E_F$. The correspondent $k$-space splitting is 0.013 \AA$^{-1}$. The FPLO procedure gives smaller splitting values: 0.0047 \AA$^{-1}$ in momentum or 20 meV in energy. Alternately switching off the hopping between different orbitals we have found that it is vertical hopping between O2$p_\sigma$ (i.e. the orbitals that form $\sigma$ bonds with Cu) that makes the main contribution to the nodal splitting value. From a simple evaluation within a 6-band model the bilayer splitting along the nodal direction ($k_x = k_y$, $0<k_x<\pi$) is $\Delta \varepsilon = 8 t_{pp} t^2_{dp} (1-\cos k_x) / \Delta^2$, where $\Delta$ = 3.4 eV is the difference in energy between the middle of the CuO conducting band and the O2$p_\sigma$ orbital, $t_{dp}$ =  1.5 eV is an in-plane Cu3$d$--O2$p$ hopping integral, and $t_{pp}$ is an effective inter-plane O2$p_\sigma$--O2$p_\sigma$ hopping integral which we estimate as 0.048 (0.02) eV within the LMTO (FPLO) scheme. This relatively large value and the orbital analysis imply that the hoppings mostly proceed via Ca atoms. The differences between LMTO and FPLO results are related to different potential construction. The larger experimental values of the BS compared with the FPLO prediction might be attributed to the mentioned different shifts of the chemical potential \cite{Drechsler}.

Since the MDC width in the vicinity of $E_F$ increases slowly with binding energy, we have traced the dispersions of both antibonding and bonding bands up to 30 meV from $E_F$ (Fig~\ref{Disp}a). Fig~\ref{Disp}b shows the width of antibonding and bonding peaks as functions of binding energy, which can be associated with the scattering rate. Such a possibility to extract the scattering rate separately for both antibonding and bonding bands could give a chance to resolve the present dilemma---what is the main boson for electrons to couple: phonons or spin fluctuations. For it has been predicted \cite{EschrigPRL2003} that the scattering by spin fluctuations should be odd (the antibonding electrons scatter to bonding band and vice versa) and should result in different widths of antibonding and bonding MDCs. Unfortunately, the differences between bonding and antibonding MDC widths presented in Fig~\ref{Disp}b stay, strictly speaking, within the experimental error and can be considered as only a hint in favor of the spin-fluctuation scenario. We also note, that the nodal splitting can be a reason for a peak-dip-hump like structure occasionally observed in the nodal direction of the cuprates which has been ascribed to strong electron-phonon interaction \cite{LanzaraNature}. The width in $k$, $\Gamma_k$, for each band at $E_F$ is about 0.012 \AA$^{-1}$ which is still 2.5 times larger than the estimated momentum resolution, $R_k$. This difference can come from roughness of the sample surface and from a finite scattering on impurities.

The significance of the O2$p_\sigma$--O2$p_\sigma$ hopping which we observe should be taken into account also in the antinodal region in order to extract  the Cu$4s$ admixture and to compare with theory \cite{LiechtensteinPRB1996}. The experimental determination of the precise amount of the admixture of such "nonstandard orbitals", being important for some aspects of the low energy-physics of various cuprates \cite{remark2},  is a challenging current problem for a future fully microscopic theory.  

In conclusion, we have experimentally resolved the bilayer splitting in the nodal direction of Bi-2212. It is observed in a wide doping range and its size, within the experimental uncertainty, does not depend on doping concentration. The value of splitting derived from the experiment is in agreement with LDA band structure calculations. This implies evidence for the lack of any electronic confinement to single planes within a bilayer in Bi-2212 due to strong correlations. The LDA orbital analysis enables us to assign the observed BS predominantly to vertical inter-plane hopping between O2$p_\sigma$ orbitals.  

We acknowledge useful discussions with H. Eschrig, K. Koepernik and T. Mishonov. We are grateful to O.~Rader for the help with the experimental setup. The project is part of the Forschergruppe FOR538 and is supported by the DFG under grants number KN393/4 and 436UKR17/10/04, and by the Fonds National Suisse de la Recherche Scientifique.


\begin{thebibliography}{}
\bibitem{bonding} We call "bonding" the lower split band at $E_F$ although, according to the band structure calculations \cite{Yaresko, Rosner}, the two split bands cross each other at about $-0.5$ eV in the nodal direction, and it is not yet clear which one has an appropriate symmetry of the wave function.
\bibitem{remark2} There are several strong hints from other experiments such as EELS \cite{moskvin}, NMR \cite{meier}, and Raman \cite{simon} and other cuprates for the presence of such nonstandard orbitals (in addition to the standard Cu 3$d_{x^2-y^2}$ and the $\sigma$ bonded O $2p$ states involved in the low-energy physics.
\end{thebibliography}
\end{document}